\documentclass[10pt,conference]{IEEEtran}
\IEEEoverridecommandlockouts
\usepackage{cite}
\usepackage{amsmath,amssymb,amsfonts}
\usepackage{algorithmic}
\usepackage{graphicx}
\usepackage{textcomp}
\usepackage{subfig}
\usepackage{braket}
\usepackage[dvipsnames]{xcolor}

\DeclareMathOperator*{\argmax}{arg\,max}
\newcommand*{\Prob}{\mathsf{Pr}}

\newcommand{\Gate}[1]{\textsc{#1}}
\newcommand{\hgate}{\Gate{h}}
\newcommand{\zgate}{\Gate{z}}

\newcommand{\xgate}{\Gate{x}}

\newcommand{\idgate}{\Gate{i}}

\newcommand{\cnotgate}{\Gate{cnot}}

\newcommand{\swapgate}{\Gate{swap}}

\def\BibTeX{{\rm B\kern-.05em{\sc i\kern-.025em b}\kern-.08em
    T\kern-.1667em\lower.7ex\hbox{E}\kern-.125emX}}
    
\input{Qcircuit}    
    
\begin{document}

\title{Error Mitigation for Deep Quantum Optimization Circuits by Leveraging Problem Symmetries\\
\thanks{This work was supported in part by the U.S.\ Department of Energy (DOE), Office of Science, Office of Advanced Scientific Computing Research AIDE-QC and FAR-QC projects and by the Argonne LDRD program under contract number DE-AC02-06CH11357. This research used resources of the Oak Ridge Leadership Computing Facility, which is a DOE Office of Science User Facility supported under Contract DE-AC05-00OR22725.}
}

\author{\IEEEauthorblockN{Ruslan Shaydulin\IEEEauthorrefmark{1} and Alexey Galda\IEEEauthorrefmark{2}\IEEEauthorrefmark{3}}
\IEEEauthorblockA{\IEEEauthorrefmark{1}Mathematics and Computer Science Division, Argonne National Laboratory, Lemont, IL 60439 USA}
\IEEEauthorblockA{\IEEEauthorrefmark{2} James Franck Institute, University of Chicago, Chicago, IL 60637 USA}
\IEEEauthorblockA{\IEEEauthorrefmark{3}Computational Science Division, Argonne National Laboratory, Lemont, IL 60439 USA}
Email: \IEEEauthorrefmark{1}rshaydulin@anl.gov, \IEEEauthorrefmark{2}agalda@uchicago.edu}

\maketitle

\begin{abstract}
High error rates and limited fidelity of quantum gates in near-term quantum devices are the central obstacles to successful execution of the Quantum Approximate Optimization Algorithm (QAOA). In this paper we introduce an application-specific approach for mitigating the errors in QAOA evolution by leveraging the symmetries present in the classical objective function to be optimized. Specifically, the QAOA state is projected into the symmetry-restricted subspace, with projection being performed either at the end of the circuit or throughout the evolution. Our approach improves the fidelity of the QAOA state, thereby increasing both the accuracy of the sample estimate of the QAOA objective and the probability of sampling the binary string corresponding to that objective value. We demonstrate the efficacy of the proposed methods on QAOA applied to the MaxCut problem, although our methods are general and apply to any objective function with symmetries, as well as to the generalization of QAOA with alternative mixers. We experimentally verify the proposed methods on an IBM Quantum processor, utilizing up to 5 qubits. When leveraging a global bit-flip symmetry, our approach leads to a 23\% average improvement in quantum state fidelity.
\end{abstract}


\section{Introduction}

The Quantum Approximate Optimization Algorithm (QAOA)~\cite{Hogg2000quantumsearch, Hogg2000, farhi2014quantum, hadfield2017quantum} is an algorithm for approximately solving combinatorial optimization problems on quantum computers. The goal of QAOA is to prepare a quantum state such that upon measuring it, a binary string that maximizes the value of the objective function is obtained with high probability. QAOA is widely considered to be one of the leading candidates for demonstrating quantum advantage on near-term hardware~\cite{Alexeev2021}. One of the main obstacles to its successful realization is the high error or noise rates of the near-term devices that lack full error correction. The limitations to QAOA success arising from noise are well documented, both from the theoretical perspective by showing bounds imposed on QAOA performance by noise~\cite{2009.05532} and from the experimental perspective by observing decreasing performance as circuit depth is increased~\cite{Lacroix2020,googleqaoaexperimental,Pagano2020}. 

\begin{figure}[t!]
    \centering
    \includegraphics[width=\linewidth]{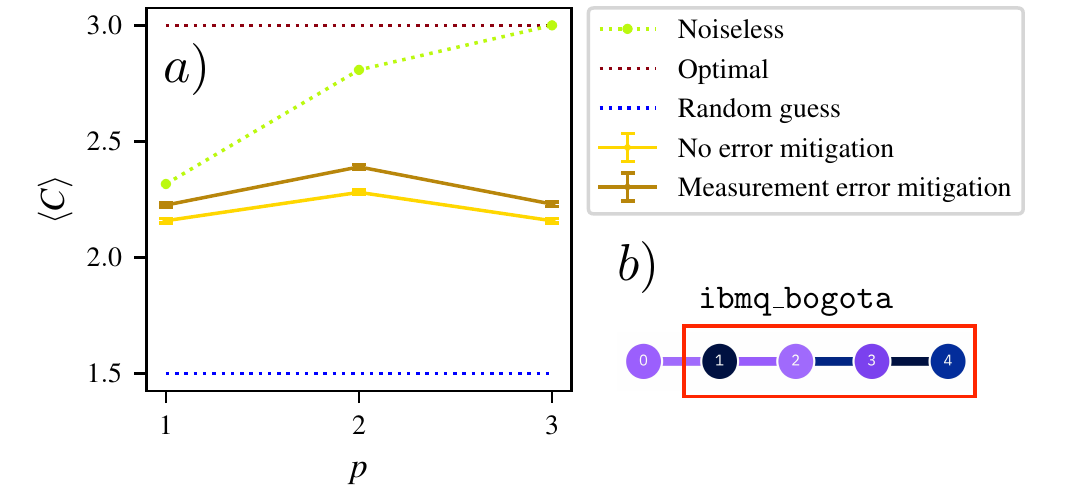}
    \caption{(a) Expected MaxCut value $\langle C\rangle$ for a binary string sampled from the QAOA state for a star graph with 4 nodes. Green represents noiseless simulation. Yellow and gold lines represent results on qubits 1--4 of \texttt{ibmq\_bogota}, which is a 5-qubit IBM Quantum Falcon processor with linear topology (b). Implementing QAOA for MaxCut on a star graph (shown in Figure~\ref{fig:graphs_jakarta}) on this device requires introducing a SWAP gate, for a total cost of 9 CNOTs per QAOA layer. We use the true optimal QAOA parameters from the dataset described in \cite{lotshaw2021empirical}. Measurement errors are mitigated by computing and inverting a calibration matrix using an open-source implementation in Qiskit~\cite{Qiskit}. Error bars show the sample mean plus or minus one standard deviation. In the absence of advanced error mitigation, the objective does not grow as $p$ is increased beyond 2.}
    \label{fig:star_4_obj_vs_p}
\end{figure}

In QAOA, pairs of alternating operators are applied in sequence, with each pair commonly referred to as a QAOA layer. On a fault-tolerant quantum processor, adding an additional layer can only increase the expected value of the objective~\cite{farhi2014quantum}. In the presence of noise and imperfect quantum gates, however, an optimal number of layers usually exists, and introducing additional layers diminishes the result. Therefore, noise and quantum gate infidelities impose an effective upper limit on the number of QAOA steps that can be executed. Figure~\ref{fig:star_4_obj_vs_p} shows an example of this behavior. In this example, increasing the number of layers $p$ beyond 2 does not lead to increased expected value of the objective when executed on a real quantum computer. At the same time, QAOA depth has to grow at least logarithmically with the problem size to allow the possibility of quantum advantage~\cite{braviyobstacles,farhi2020qaoaneedsfullgraphtypical,farhi2020qaoaneedsfullgraphworst}. Increasing the number of QAOA layers that can be executed before reaching the limit imposed by noise is therefore essential to the successful application of QAOA. This can be achieved either by reducing the error rates of the hardware or by mitigating the errors through software or algorithmic techniques.

A particularly promising direction for mitigating the errors and improving the performance of near-term algorithms on noisy quantum hardware consists in leveraging the specific structure of the circuit being executed. One such approach is the recently introduced error mitigation by symmetry verification, proposed in \cite{BonetMonroig2018,McArdle2019}
and experimentally demonstrated using two transmon qubits in \cite{Sagastizabal2019}. This approach uses the observation that many near-term algorithms preserve certain symmetries of the system, thereby restricting the (noiseless) evolution to the corresponding eigenspace of the symmetry operators. Then if the system is pushed out of this eigenspace by an error during the execution of the quantum circuit, the error can be partially mitigated by projecting the system back into the correct eigenspace.

In this work we propose an error mitigation technique for the Quantum Approximate Optimization Algorithm by verifying the classical symmetries of the objective function preserved by the QAOA ansatz. Specifically, we show that by projecting into the $+1$ eigenspace of the corresponding symmetry operators, we can improve the fidelity of the quantum state, leading to a higher expected objective value and increased probability of  success. We demonstrate the efficacy of the proposed approach by applying it to QAOA circuits for the MaxCut problem on four graphs with up to four nodes. We show improvement in quantum state fidelity, expected objective function value, and the probability of sampling the optimal solution on \texttt{ibmq\_jakarta}, which is a cloud-based IBM Quantum processor~\cite{ibmq_systems}, using up to 5 qubits. We complement the results obtained on hardware with simulations, further highlighting the promise of the proposed approach on future devices with reduced noise rates.

Our results improve on the previous state of the art in the following ways. We extend the symmetry verification approach proposed for quantum simulation in \cite{BonetMonroig2018,McArdle2019} to optimization problems. Specifically, we focus on the classical symmetries of the objective function present in the QAOA ansatz. We leverage classical variable (qubit) index permutation symmetries, which were not considered in the previous works~\cite{BonetMonroig2018,McArdle2019,Streif2021}. We show extensive experimental verification on quantum hardware utilizing up to 5 qubits of a 7-qubit superconducting device. To the best of out knowledge, this is the first work to advocate an application-specific error mitigation technique in gate-model quantum optimization that has been validated in experiments on hardware. 

\section{Background}

In this section we introduce the notation used in this paper and briefly review our notions of classical and quantum optimization and the relevant prior results on the role of symmetry in QAOA and error mitigation by symmetry verification.

\subsection{Binary optimization}

Consider a real-valued objective function $f$ defined on the Boolean cube and the corresponding optimization problem
\[
\max_{x\in \{0,1\}^n} f(x).
\]

For clarity of presentation, we restrict ourselves to maximization problems, though all results in this work can be analogously applied to minimization. We denote the binary string that maximizes $f(x)$ as 
\[
x^* = \argmax_{x\in \{0,1\}^n} f(x).
\]

This binary string need not be unique. The objective function $f$ can be encoded as an $n$-qubit Hamiltonian $C$ that acts on computational basis states as $C\ket{x} = f(x)\ket{x}$, $x\in \{0,1\}^n$, with compact representations in terms of Pauli $\zgate$s available for many relevant functions~\cite{hadfieldrepresentation}.

\subsection{MaxCut}

Consider an undirected graph $G$ with a set of vertices $V$ and a set of edges $E$. The objective of MaxCut is to find an assignment of vertices into two disjoint subsets such that the number of edges spanning both subsets is maximized. The MaxCut objective is encoded by the Hamiltonian

\[
C_{\text{MaxCut}} = \frac{1}{2}\sum_{(j,k)\in E}(\idgate-\zgate_j\zgate_k).
\]

MaxCut on general graphs is APX-complete~\cite{papadimitriou1991optimization}.

\subsection{Quantum Approximate Optimization Algorithm}

The Quantum Approximate Optimization Algorithm  is a hybrid quantum-classical algorithm to approximately solve combinatorial optimization problems. It combines a parameterized quantum evolution with a classical ``outer-loop'' method to find good parameters. The parameterized quantum evolution is performed by applying a sequence of alternating operators
\[
\ket{\vec{\beta}, \vec{\gamma}} := U_B(\beta_p)U_C(\gamma_p)\ldots U_B(\beta_1)U_C(\gamma_1)\ket{s},
\]
where $C$ is the problem Hamiltonian, $U_C(\gamma) = e^{-i\gamma C}$ is the phase operator, $U_B(\beta)=e^{-i\beta\sum_jX_j}$ is the mixing operator, and $\ket{s}=\ket{+}^{\otimes n}$ is the initial uniform superposition product state. The evolution is followed by the measurement in the computational basis, with measurement outcome probabilities following the Born rule 
\[
\Prob(x) = |\braket{x|\vec{\beta},\vec{\gamma}}|^2, \quad x\in \{0,1\}^n.
\]

The goal of the outer-loop classical routine is to produce parameters that maximize the expected objective value of the quantum evolution measurement outcomes:
\[
\langle C\rangle := \bra{\vec{\beta}, \vec{\gamma}}C\ket{\vec{\beta}, \vec{\gamma}} = \sum_{x\in \{0,1\}^n}\Prob(x)f(x).
\]

When executing QAOA on quantum hardware, we do not have access to the true value of $\langle C\rangle$. In those cases, we will slightly abuse the notation and use $\langle C\rangle$ to denote the sample mean $\sum_{i=1}^n f(x_i) / n$ obtained from the measurement outcomes. One can easily see that in the absence of noise and in the limit $n\rightarrow \infty$, the two values are equal. 

\subsection{Symmetries in QAOA}

The QAOA ansatz $\ket{\vec{\beta}, \vec{\gamma}}$ is invariant under a symmetry $S$ if it commutes with both the problem and the mixing Hamiltonians (i.e.,  $[S,C] = 0 = [S,B]$) and the QAOA initial state is invariant under it (i.e., $S\ket{s}=\ket{s}$)~\cite{shaydulinsymm}. The specific class of symmetries of the QAOA ansatz considered in this paper is the symmetries of the objective function. A permutation $a\in S_{2^n}: x\rightarrow a(x)$ of the $n$-bit strings is a symmetry of an objective function $f$ on $n$ bits if it leaves the objective function unchanged; that is, $f(x) = f(a(x)), \forall x\in \{0,1\}^n$. For each $a\in S_{2^n}$, its representation on qubits acting as $A\ket{x}=\ket{a(x)}, \forall x\in \{0,1\}^n$, can be constructed as follows:
\[
A =\sum_{x\in\{0,1\}^n}\ket{a(x)_1\ldots a(x)_{n}}\bra{x_1\ldots x_n}.
\]

An important subgroup is formed by permutations that are variable index permutations, that is, $a\in S_n$. If $a\in S_n$ is the symmetry of the objective function, then its representation on qubits $A$ is a symmetry of the QAOA ansatz~\cite{shaydulinsymm}. While a sufficiently large symmetry group may reduce the effective dimensionality of the solution space and the Hilbert space in which QAOA dynamics take place from exponential to polynomial in number of variables, this does not necessarily allow one to simulate the QAOA circuits or to produce solutions for the optimization problem~\cite[Sec 3.2]{shaydulinsymm}. In other words, in general a problem with a lot of symmetry can still be hard to solve and the corresponding QAOA circuits hard to classically simulate. For a detailed discussion of the role of classical symmetries in the QAOA ansatz and their applications, the reader is referred to~\cite{shaydulinsymm,Shaydulin2021symmtrain}.

\subsection{Error mitigation by symmetry verification}\label{sec:em_sv_background}

A unitary operator $S$ is a symmetry of an operator $Q$ if $Q$ is invariant under $S$, that  is, if $S^\dagger QS = Q$, or, equivalently, $[Q,S]=0$. The study of the eigenstate of a Hamiltonian $H$ with symmetry $S$ can be performed entirely within a single target eigenspace $\mathcal{S}$ of $S$. Then if an error pushes the system outside of $\mathcal{S}$ during the execution on a noisy quantum computer, this error can be mitigated by verifying the symmetry and discarding the results where the symmetry is not preserved.

Error mitigation is performed by measuring the value of the operator $S$ and postselecting on the measurement outcomes. This is done by performing a projector-valued measurement $\{M_i\}$, where $M_s$ is the projector onto the eigenspace corresponding to the eigenvalue $s$ of $S$. If the evolution is restricted to the eigenspace of $S$ corresponding to the eigenvalue $s$ (which is the case when studying a particular eigenspace of a Hamiltonian with symmetry $S$), then, in the absence of errors, $M_s\ket{\psi}=\ket{\psi}$ at any point throughout the evolution. That is, in the absence of errors, the measurement $M_s$ does not change the quantum state. Here $\ket{\psi}$ denotes the true noiseless state, e.g. $\ket{\psi}= \ket{\vec{\beta}, \vec{\gamma}}$ if the projection is performed at the end of the QAOA evolution. If some error is present, performing the projector-valued measurement $M_s$ and postselecting on the outcomes produce a state $\rho_s$ with a greater or equal overlap with the target (noiseless) state~\cite{BonetMonroig2018}:
\begin{multline}
\text{Tr}[\rho_s\ket{\psi}\bra{\psi}] = \frac{\text{Tr}[M_s\rho M_s\ket{\psi}\bra{\psi}]}{\text{Tr}[M_s\rho]} = \frac{\text{Tr}[\rho \ket{\psi}\bra{\psi}]}{\text{Tr}[M_s\rho]}\geq \\ \geq \text{Tr}[\rho\ket{\psi}\bra{\psi}],
\end{multline}
where $\rho$ is the density matrix before symmetry verification. As the measurement being performed is non-destructive, the procedure can be applied at the end or in the middle of the quantum circuit and can be extended to multiple commuting symmetry operators by performing measurement and postselection in sequence.

The symmetries we consider in this paper are operators with eigenvalues $s=\pm 1$. In this case the projectors can be written as $M_{s} = \frac{1}{2}(\idgate\pm S)$. The projection into the $+1$ eigenspace of $S$ can be performed by applying the following circuit and performing postselection on the measurement outcome on the ancilla being equal to $0$.

\begin{equation}
  \begin{gathered}
    \Qcircuit @C=1em @R=1em {
\lstick{\ket{\psi_{\text{in}}(\theta)}}     & \qw       & \gate{S}      &  \qw   &  \qw & \rstick{\ket{\psi_{\text{out}}(\theta)}} \\
\lstick{\ket{0}}            & \gate{\hgate} & \ctrl{-1} & \gate{\hgate} & \meter & \cw
}
\end{gathered}
\label{eq:projector_circ}
\end{equation}

This symmetry verification approach can be understood as a restricted version of the stabilizer parity
measurements used, for example, in surface codes~\cite{Fowler2012}. In the case of surface
codes, the logical qubit is created by forcing the evolution to be restricted
to a simultaneous eigenstate of the stabilizer operators; if an error occurs,
stabilizer parity measurement either corrects the error by projecting the
system back into the simultaneous eigenstate or produces a measurement result
that can be used to correct the error in postprocessing. In the case of
error mitigation by symmetry verification, the constraints imposed by the
symmetries of the system are much looser, and enforcing them incurs
a correspondingly lower overhead. The parity checks (symmetries) that are being measured are insufficient to ``narrow down'' (detect) which error has happened; rather, the method relies on the symmetry measurements projecting the state into the correct subspace and on postselection to discard the results where the check failed.

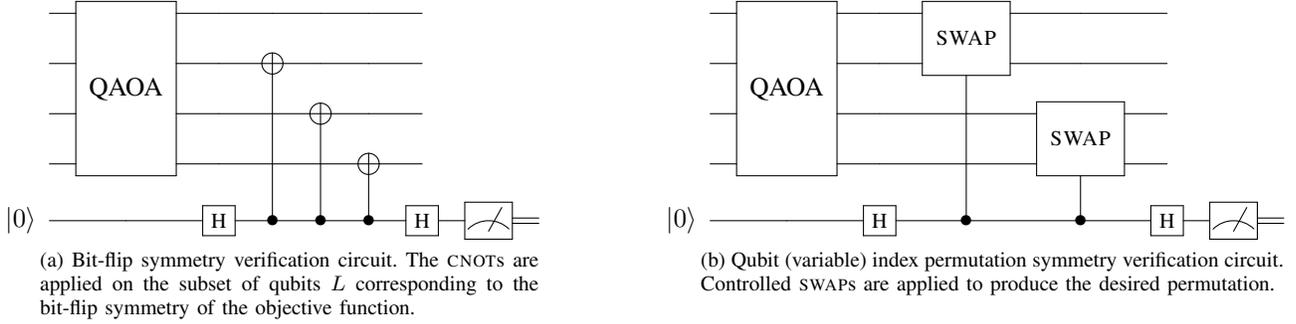
\begin{figure*}[ht!]
    \hspace{0.45in}
    \subfloat[Bit-flip symmetry verification circuit. The $\cnotgate$s are applied on the subset of qubits $L$ corresponding to the bit-flip symmetry of the objective function.\label{fig:xxx_projector}]{
    \Qcircuit @C=1em @R=1em{
        & \multigate{3}{\text{QAOA}} & \qw & \qw     & \qw   & \qw   & \qw \\
        & \ghost{\text{QAOA}}        & \qw & \targ     & \qw   & \qw   & \qw \\
        & \ghost{\text{QAOA}}        & \qw & \qw       & \targ & \qw   & \qw\\
        & \ghost{\text{QAOA}}        & \qw & \qw       & \qw   & \targ & \qw\\
        \lstick{\ket{0}} & \qw        & \gate{\hgate} & \ctrl{-3} & \ctrl{-2} & \ctrl{-1} & \gate{\hgate} & \meter & \cw
}}
\hspace{0.8in}
    \subfloat[Qubit (variable) index permutation symmetry verification circuit. Controlled $\swapgate$s are applied to produce the desired permutation.\label{fig:xy_projector}]{
    \Qcircuit @C=1em @R=1em{
        & \multigate{3}{\text{QAOA}} & \qw & \multigate{1}{\swapgate}    & \qw   & \qw   & \\
        & \ghost{\text{QAOA}}        & \qw & \ghost{\swapgate}  & \qw    & \qw \\
        & \ghost{\text{QAOA}}        & \qw &  \qw       & \multigate{1}{\swapgate}   & \qw\\
        & \ghost{\text{QAOA}}        & \qw & \qw       & \ghost{\swapgate}   & \qw\\
        \lstick{\ket{0}} & \qw        & \gate{\hgate} & \ctrl{-3} & \ctrl{-1} &  \gate{\hgate} & \meter & \cw
}}
    \caption{Circuits used for verification of objective function symmetries. In order to perform error mitigation, the measurement outcomes are postselected on the ancilla being equal to $0$. The procedure can be applied at the end of the circuit (pictured), as well as inserted between the alternating operators throughout the evolution.}
    \label{fig:verification_circuits}
\end{figure*}

\section{Verification of the objective function symmetries}\label{sec:verification_theory}

We mitigate the errors that occur during QAOA evolution executed on a noisy quantum computer by verifying the classical symmetries of the objective function. The verification is performed by projecting into the $+1$ eigenspace of the corresponding symmetry operators. Specifically, we consider two classes of objective function symmetries: bit-flip symmetries and variable (qubit) index permutation symmetries.

The first class of symmetries we consider is the bit-flip symmetries. These are symmetries of the form $a(x) = x\oplus l$, where $l$ is a binary string with $l_j = 1, \forall j\in L$, $f_j=0$ otherwise, and $L$ is the subset of bits being flipped. A common example is a global bit-flip symmetry with $L = \{1,\ldots,n\}$. Bit-flip symmetries can be represented on qubits as $A=\bigotimes_{j\in L}\xgate_j$, where $\xgate_j$ is the Pauli $\xgate$ on qubit $j$. 

If $a$ is a symmetry of the objective function, then $[A,C]=0$. Since $[A,B]=0$ and $A\ket{s}=\ket{s}$ trivially, $A$ is a symmetry of the QAOA ansatz. Since $A$ has eigenvalues $\pm 1$, the projector circuit can be obtained by using the construction in~(\ref{eq:projector_circ}) as shown in Fig.~\ref{fig:xxx_projector}, where $\cnotgate$s are applied on the subset of qubits $F$. Then the verification can be performed by applying the projector circuit and postselecting the measurement outcomes on the ancilla being equal to 0 (corresponding to the eigenvalue $+1$ of $A$). Note that the verification need not be performed at the end of the circuit but can be performed intermittently throughout the circuit. 

The second class of  objective function  symmetries we consider is the variable (qubit) index permutation symmetries, that is, symmetries $a\in S_n$. If $A$ is the corresponding qubit permutation representing $a$, it commutes with the mixing Hamiltonian since it acts uniformly on all qubits. Moreover, $A\ket{s}=\ket{s}$ trivially; therefore $A$ is a symmetry of the QAOA ansatz. Permutation symmetries can be represented on qubits as a combination of $\swapgate$ operators.

For simplicity, in the experiments presented in Sec.~\ref{sec:experiments}, we consider only the subset of permutation symmetries that can be represented by a sequence of non-overlapping $\swapgate$s. In this case the symmetry operator has eigenvalues $\pm 1$, and therefore the symmetry verification circuit can be constructed following~(\ref{eq:projector_circ}), as shown in Fig.~\ref{fig:xy_projector}, where $\swapgate$s are applied to produce the desired permutation. The error mitigation is performed in the same way as in the bit-flip case. We note that both symmetries can be verified simultaneously as the corresponding operators commute with each other.

As discussed in Sec.~\ref{sec:em_sv_background}, if no errors occur during the execution of the verification circuit, the proposed procedure can only improve the overlap between the noisy quantum state after the verification and the true noiseless state of the system. Since the goal of QAOA is not to prepare a particular target state (as is the case in quantum simulation) but to solve a classical optimization problem, two important figures of merit to consider are the expected value of the objective function $\langle C\rangle$ and the probability $\Prob(x^*)$ of sampling a binary string that maximizes the objective function. The two values are closely connected because of the distribution of the objective function values of QAOA samples being concentrated around the mean~\cite{farhi2014quantum}. While the verification procedure is guaranteed to increase the overlap with the noiseless state, it is not guaranteed to improve $\langle C\rangle$ or $\Prob(x^*)$ in general. In fact, improving the overlap (or fidelity) can decrease $\langle C\rangle$ if the noise has pushed the state closer to the solution.

In practice, the verification circuit is executed on noisy hardware and may introduce additional errors. For the symmetry verification procedure to be successful, the overhead of additional errors should not outweigh the benefits from error mitigation. Therefore it is essential to compile the circuits in Fig.~\ref{fig:verification_circuits} as efficiently as possible. The verification circuits considered in this work implement Clifford unitaries, thus making it feasible to compile them optimally for up to 6 qubits~\cite{optimal6qubit} and significantly reduce the $\cnotgate$ count for circuits on larger numbers of qubits~\cite{bravyicliffordopt}. Compiling these unitaries for nearest-neighbor architectures such as the superconducting qubits employed in IBM Quantum devices~\cite{ibmq_systems} can be challenging in practice because of the need to minimize the number of $\swapgate$s introduced. However, the per-instance cost of optimizing the circuits can be significantly reduced by noticing that for a particular device topology (e.g., linear) and a particular symmetry (e.g., global bit-flip symmetry), the same verification circuit is applied irrespective of the objective function exhibiting the symmetry. Therefore the circuit can be optimized once and stored in a database to be used in QAOA runs on different problems that share the same symmetry. In this way, the high cost of finding an optimized implementation of the symmetry verification unitary can be amortized over many QAOA runs.

\subsection{Application to QAOA for MaxCut}

\begin{figure*}
    \centering
    \includegraphics[width=0.95\textwidth]{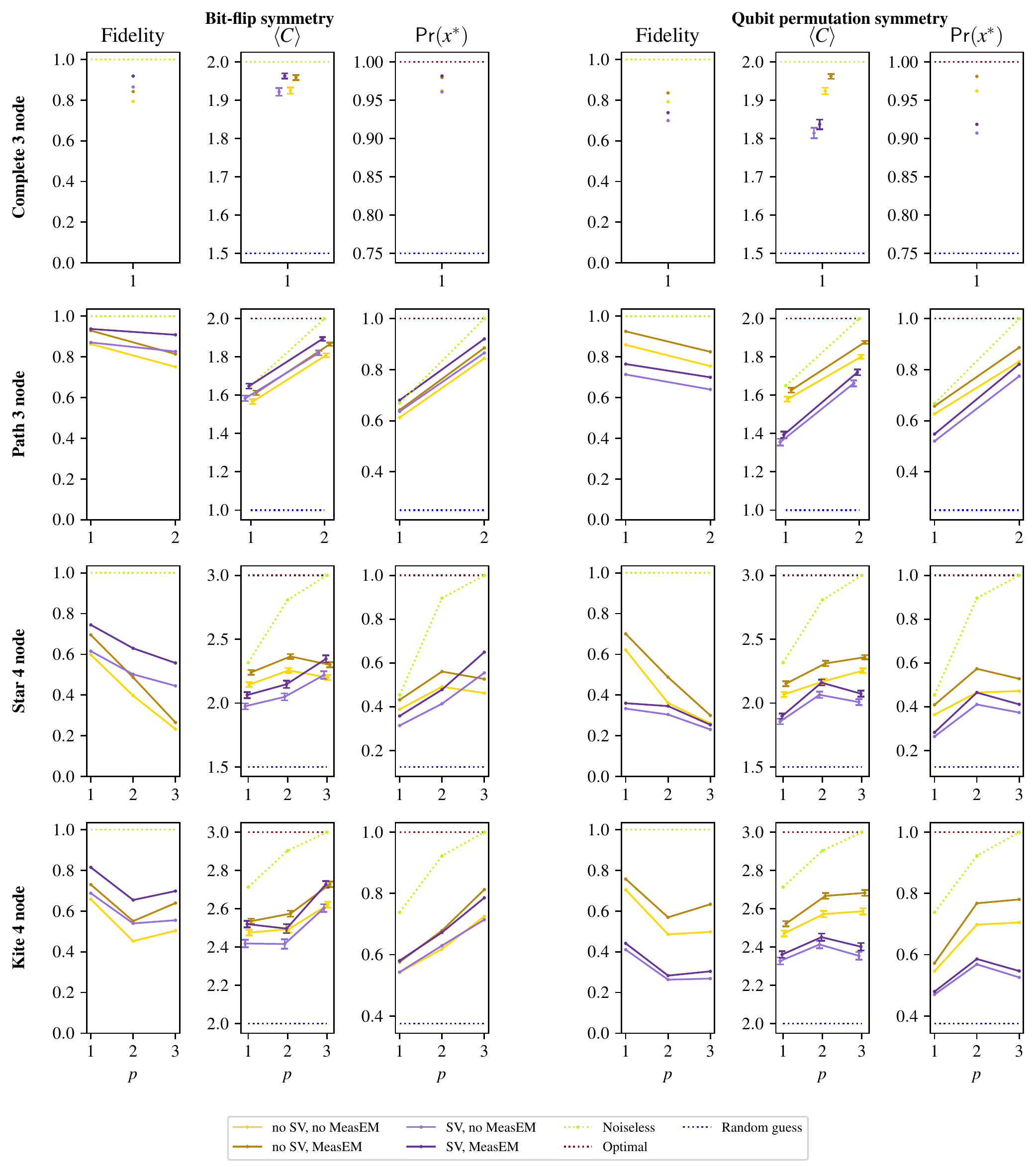}
    \caption{Symmetry verification (SV) for QAOA for MaxCut on \texttt{ibmq\_jakarta}. MeasEM denotes the results with measurement error mitigation, which is performed by computing and inverting a calibration matrix using an open-source implementation in Qiskit~\cite{Qiskit}. Error bars on the value of $\langle C\rangle$ show the sample mean plus or minus one standard deviation. We observe fidelity improvement over baseline (no SV) from symmetry verification for all instances with bit-flip symmetry.}
    \label{fig:jakarta_res}
\end{figure*}

\begin{figure}[t]
    \centering
    \includegraphics[width=0.95\linewidth]{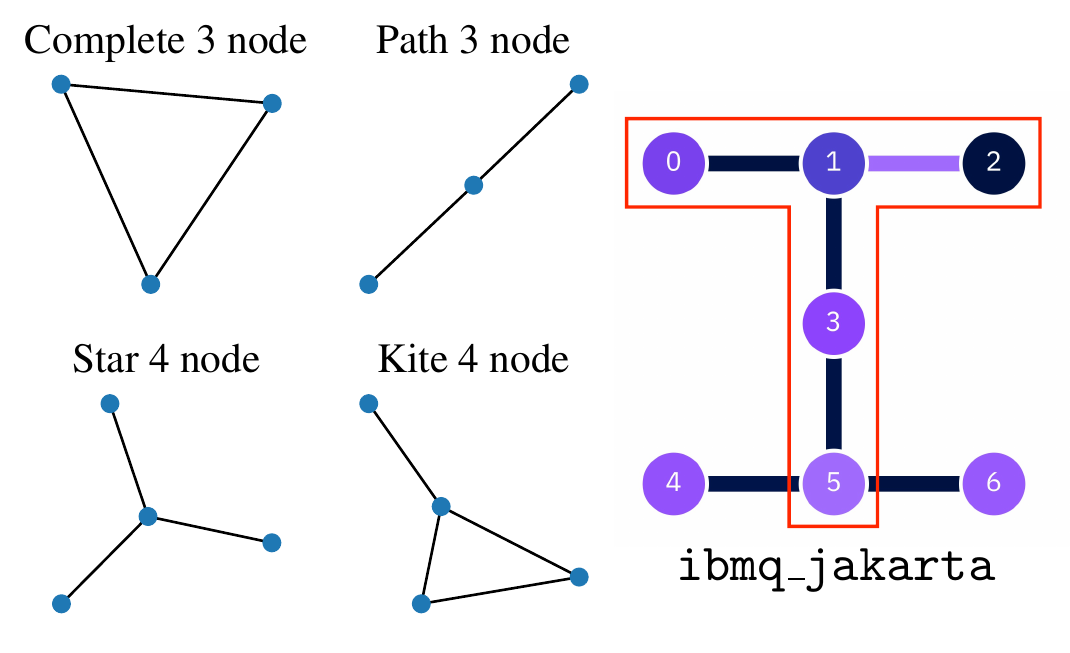}
    \caption{Graphs for the MaxCut instances considered and the \texttt{ibmq\_jakarta} layout. The qubits used in experiments are highlighted in red.}
    \label{fig:graphs_jakarta}
\end{figure}

For QAOA applied to the MaxCut problem, the first symmetry we consider is a global bit-flip symmetry ($\mathbb{Z}_2$ symmetry of the problem Hamiltonian) exhibited by all MaxCut instances. The second class of symmetries is  variable (qubit) index permutation symmetries that are the symmetries of the underlying graph~\cite{shaydulinsymm} and are specific to a particular problem instance. The graph symmetries we consider are (vertex) automorphisms, that is, permutations of nodes that preserve edges.  In general, no efficient algorithm is known for computing the automorphism group of a graph. However, the automorphism group for a particular graph instance can be constructed efficiently for some problem classes such as bounded-degree graphs~\cite{luks1982isomorphism} and in practice can be computed by using fast solvers with worst-case exponential running time in seconds for graphs with up to tens of thousands of nodes~\cite{McKay201494,Shaydulin2021symmtrain}.

\section{Experimental results}\label{sec:experiments}

\begin{figure*}[htb]
    \centering
    \includegraphics[width=0.95\textwidth]{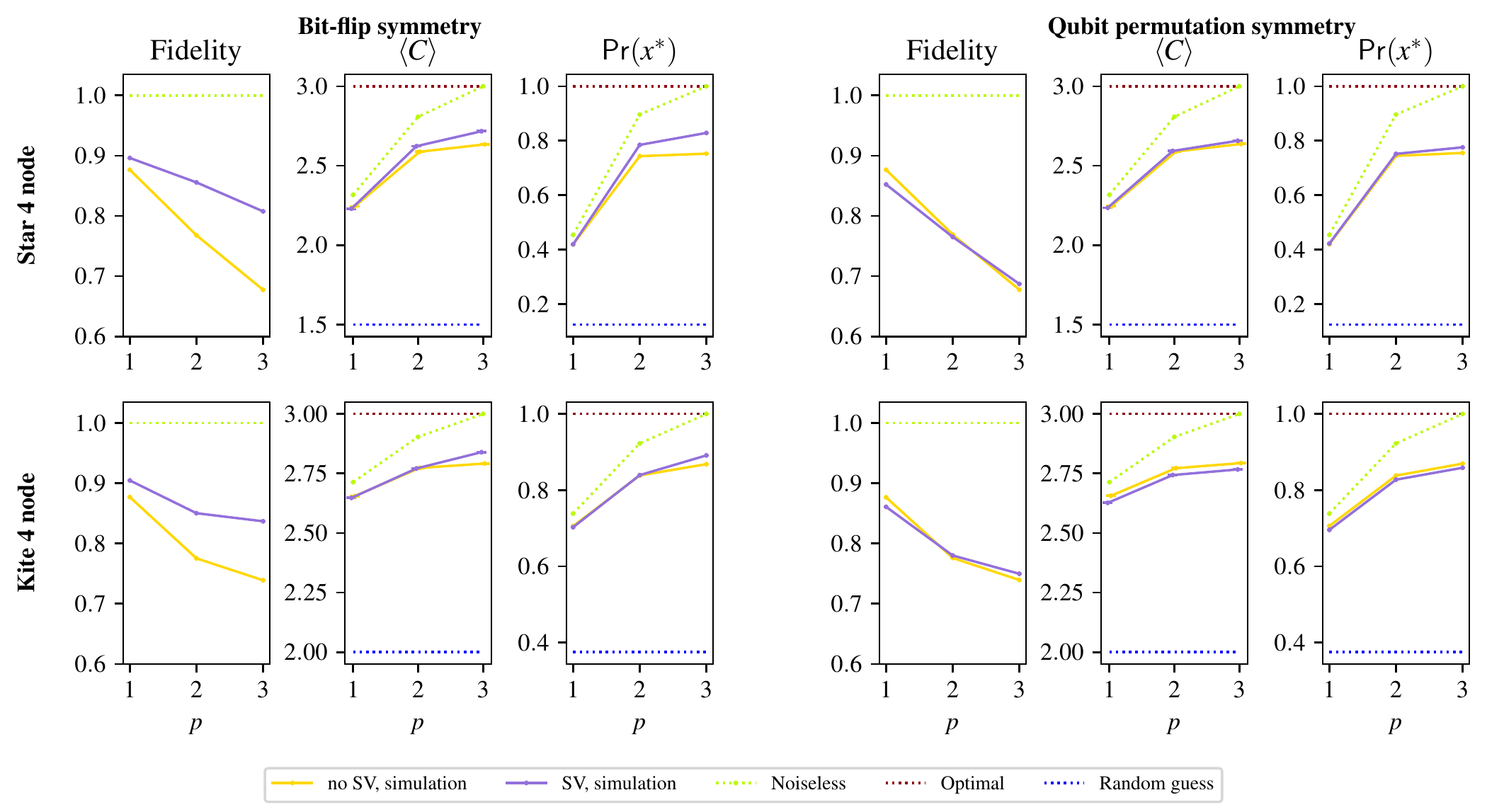}
    \caption{Symmetry verification (SV) results obtained in simulation with depolarizing noise channel and thermal relaxation. The error model is calibrated using the parameters obtained from \texttt{ibmq\_jakarta} calibration data. We observe fidelity improvement from qubit permutation symmetry verification for both graphs with $p=3$.}
    \label{fig:sim_res}
\end{figure*}

\begin{figure}[t]
    \centering
    \includegraphics[width=0.95\linewidth]{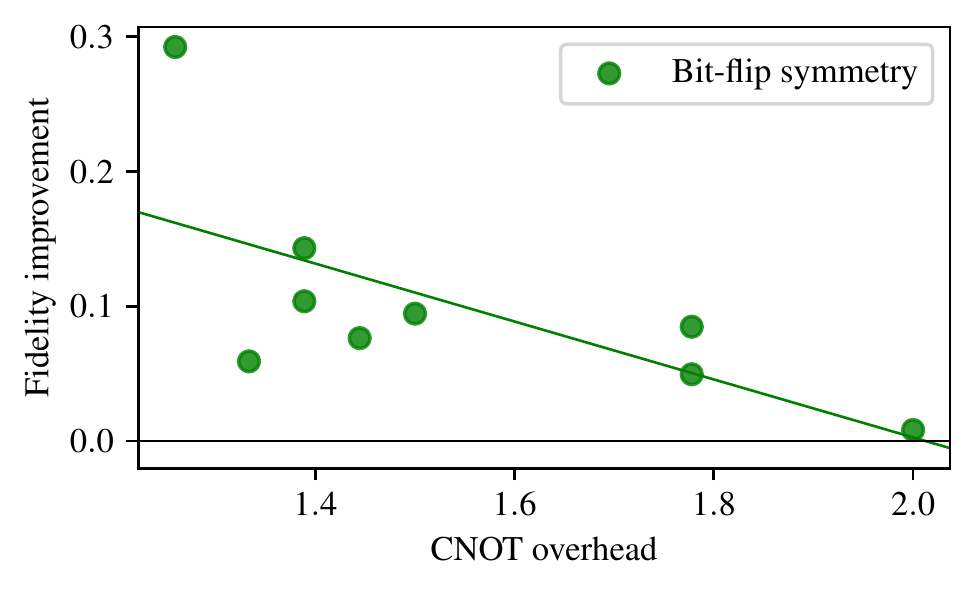}
    \caption{The improvement in fidelity from error mitigation as the function of the relative $\cnotgate$ count overhead for bit-flip symmetry verification on \texttt{ibmq\_jakarta}. The improvement in fidelity is the difference between fidelity with and without error mitigation. The relative $\cnotgate$ overhead is the ratio between the number of $\cnotgate$s in the QAOA circuit with and without symmetry verification. As the overhead decreases, fidelity improvement increases. The solid line shows a linear regression, which we include for visualization purposes. See Figure~\ref{fig:jakarta_res} for additional details.}
    \label{fig:cnot_overhead}
\end{figure}

To demonstrate the efficacy of the proposed methods, we apply them to QAOA circuits used to solve four MaxCut instances on graphs with 3 and 4 nodes. We perform experiments utilizing up to 5 qubits on a 7-qubit \texttt{ibmq\_jakarta} superconducting processor, which is one of the IBM Quantum Falcon processors~\cite{ibmq_systems}. We have made all the quantum circuits used in the experiments available online~\cite{codeanddata}.

We use QAOA to solve MaxCut on all non-isomorphic connected 3-node graphs and on two 4-node graphs. We consider all values of the number of layers $p$ up to $p_{\max}$ at which QAOA solves the problem exactly. As QAOA with $p_{\max}$ layers solves the problem exactly, adding extra layers beyond $p_{\max}$ does not change the final quantum state and is therefore unnecessary. Specifically, we consider the following four graphs with symmetry: 3-node path graph ($p_{\max}=1$), 3-node complete graph ($p_{\max}=2$), 4-node star graph ($p_{\max}=3$), and 4-node kite graph ($p_{\max}=3$). In all experiments we map the QAOA circuits on qubits 0,1,3,5 of \texttt{ibmq\_jakarta} and use qubit 2 as the ancilla. We choose these qubits because of their lower observed error rates and higher observed gate fidelities. The graphs and the device layout are presented in Fig.~\ref{fig:graphs_jakarta}.

We perform verification of two types of symmetries:  the global bit-flip symmetry exhibited by any MaxCut problem and  local graph symmetries. To reduce the overhead of error mitigation in terms of number of gates, in this paper we  consider only local permutation symmetries for 3-node complete and 4-node star and kite graphs that are implemented by one $\swapgate$. For the 3-node path graph we consider a mirror symmetry that is also implemented by one $\swapgate$. We apply symmetry verification at the end of QAOA evolution, although the verification can also be performed throughout the QAOA evolution.

To avoid the need to optimize QAOA parameters and to guarantee that the parameters we obtain are indeed optimal, we use parameters from the dataset described in \cite{lotshaw2021empirical}. Since parameters can be frequently found purely classically~\cite{yuriQAOAsim,Shaydulin2021symmtrain} or predicted by using a machine learning model~\cite{khairy2019learning,verdon2019learning,wilson2019optimizing}, executing QAOA on a quantum computer only to sample the solutions is a practical relevant setting. While we do not evaluate the effects of the proposed methods on the performance of the parameter optimization methods, we expect that the improved quantum state fidelity would make the parameter optimization easier. We show that our methods lead to more accurate sample estimates of the expectation $\langle C\rangle$, which is commonly used by outer-loop classical optimizers~\cite{Shaydulin2019MultistartDOI}. Additionally, we expect the proposed methods to improve the accuracy of other statistics used by optimizers, such as conditional value at risk~\cite{Barkoutsos2020}.

To compile the verification circuits, we use the quantum gate synthesis algorithm QSearch~\cite{Davis2020}. The bit-flip symmetry verification circuit is the same across all instances with the same number of nodes, and the two 4-node instances in our benchmark have the same qubit permutation symmetry verification circuit, enabling us to optimize these circuits only once and use them for multiple different instances. The high-cost synthesis method we use to compile the circuits into as few $\cnotgate$s as possible does not scale to larger numbers of qubits. For larger instances, alternative scalable compilation techniques such as~\cite{bravyicliffordopt} can be used.

In Fig.~\ref{fig:jakarta_res} we present the results obtained on \texttt{ibmq\_jakarta}. Three figures of merit are shown: quantum state fidelity, expected value of the objective $\langle C\rangle$, and probability $\Prob(x^*)$ of sampling a binary string that maximizes the objective function. We additionally plot the values of the statistics in the absence of noise and their optimal values. Error bars on the value of $\langle C\rangle$ show the sample mean plus or minus one standard deviation. To compute the quantum state fidelity, we perform quantum state tomography. We use maximum likelihood estimation to infer the quantum state from the statistics of measurements, which is the most commonly used approach. We note that this method is known to be incompatible with error bars~\cite{Faist2016,BlumeKohout2010}. We discuss the results on hardware in detail in Sec.~\ref{seq:results_hardware}.

To complement the hardware results and to showcase the potential of the proposed error mitigation techniques in the regime where the hardware has lower error rates, we additionally include the results obtained in simulation using a noise model that includes only  depolarizing noise and thermal relaxation. We discuss the simulation results in Sec.~\ref{seq:results_simulation}.

\subsection{Mitigation of errors on a 7-qubit superconducting device}\label{seq:results_hardware}

Fig.~\ref{fig:jakarta_res} shows the results obtained by applying symmetry verification to QAOA circuits for MaxCut on \texttt{ibmq\_jakarta}. We observe fidelity improvement from global bit-flip symmetry verification for all instances considered (leftmost column of Fig.~\ref{fig:jakarta_res}). Note that the fidelity of the quantum state in general decreases with increasing circuit depth (here, the number of QAOA layers $p$) due to error accumulation. However, an increase in $\langle C\rangle$ and $\Prob(p^*)$ from the additional QAOA layers is still possible, as long as fidelity does not decrease too quickly. This increase comes from the fact that the values of $\langle C\rangle$ and $\Prob(p^*)$ grow with depth $p$ for the noiseless state, offsetting the decreasing fidelity between the noiseless state and the state on the hardware.

As discussed in Sec.~\ref{sec:verification_theory}, for a given depth $p$ fidelity improvement does not necessarily lead to an improvement in either the expectation of the objective $\langle C\rangle$ or the probability of sampling the optimal solution $\Prob(x^*)$. However, the closer the QAOA state is to solving the problem exactly, the more likely it is that the improvement in fidelity will lead to improvement in $\langle C\rangle$ or $\Prob(x^*)$. Performance of bit-flip symmetry verification on the 4-node star graph is an example of this behavior. While we observe fidelity improvements for all values of $p$, we  observe an improvement in $\langle C\rangle$ and $\Prob(x^*)$ only or $p=3$, when in the noiseless case the QAOA solves the problem exactly.

The proposed error mitigation methods introduce a nontrivial overhead in terms of additional gates that need to be executed on the noisy quantum computer. This overhead can be partially reduced by advanced compilation techniques that reduce the number of additional $\cnotgate$s introduced for symmetry verification. We show the relationship between the improvements in quantum state fidelity from symmetry verification and the relative overhead in $\cnotgate$ count in Fig.~\ref{fig:cnot_overhead}. We observe that as relative overhead goes down, fidelity improvement increases. This makes the proposed methods especially promising for problems with a larger number of variables, which require a higher QAOA depth $p$ to be solved exactly, leading to a lower relative overhead of error mitigation.

We do not observe improvements in fidelity from the qubit (variable) index permutation symmetry verification on \texttt{ibmq\_jakarta}. We hypothesize that this effect is due to the higher relative overhead and the permutation symmetries considered being local. To highlight the promise of this approach in the regime with fewer sources of error on hardware, we complement the results obtained on \texttt{ibmq\_jakarta} with noisy simulations, which we discuss below.

\subsection{Simulation experiments}\label{seq:results_simulation}

We present the noisy simulation results in Fig.~\ref{fig:sim_res}. Since the main goal for simulation experiments is to demonstrate the potential of the qubit permutation symmetry verification, we  include only instances that have sufficiently deep QAOA circuits for the permutation symmetry verification to be efficacious. The noisy simulations are performed  by using the Aer module of the Qiskit software framework. The noise model objects are derived by using the calibrated gate error rates and coherence times on corresponding quantum processor, and supplied to the noisy quantum circuit simulator. Single-qubit gate errors are modeled as a depolarizing channel and thermal relaxation with rates governed by the device’s calibrated $T_1$ and $T_2$ coherence times, correspondingly, and the single-qubit gate duration. The noise model also accounts for depolarizing errors of two-qubit gates according to the latest device calibration information, as provided by IBM Quantum~\cite{ibmq_systems}. In particular, we employ the density matrix simulation method that samples measurement outcomes from noisy circuits. We intentionally do not include readout errors in the noisy simulations to account for the measurement error mitigation step performed on the outcomes of experiments executed on quantum hardware.

\begin{figure}[htb!]
    \centering
    \includegraphics[width=0.95\linewidth]{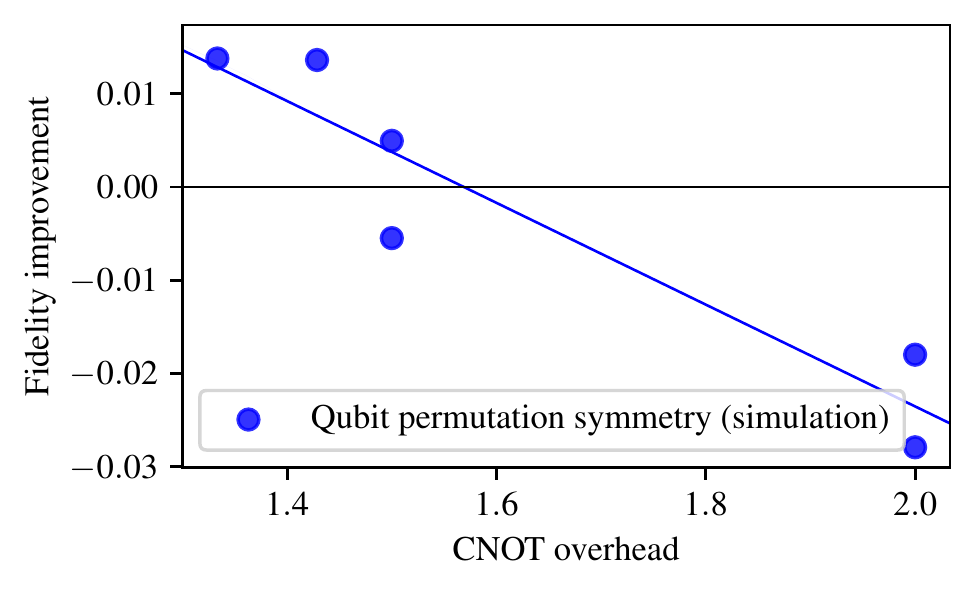}
    \caption{The improvement in fidelity from error mitigation as the function of the relative $\cnotgate$ count overhead for bit-flip symmetry verification in simulation. As the overhead decreases, fidelity improvement increases. The solid line shows a linear regression, which we include for visualization purposes. See Figure~\ref{fig:sim_res} for additional details.}
    \label{fig:cnot_overhead_sim}
\end{figure}

In agreement with the results obtained on the hardware, we observe that the improvement in fidelity grows as the relative overhead of symmetry verification decreases. Unlike for bit-flip symmetry, for qubit permutation symmetry we do not observe improvement for all values of $p$. In fact, we  observe improvement in fidelity in simulation only for $p=3$. This supports the intuition that because  of the high hardware error rates, the overhead of the permutation symmetry verification is too high to be efficacious for the relatively shallow circuits considered. Since for larger instances a higher depth $p$ is required, the relative overhead of the symmetry verification will be lower, making it beneficial on hardware. In Fig.~\ref{fig:cnot_overhead_sim} we show the relationship between the overhead and fidelity improvement for the qubit permutation symmetry obtained in simulation.

\section{Discussion}

While QAOA is a promising approach for using quantum computers to accelerate the solution of combinatorial optimization problems, it requires relatively deep circuits in order to be successful. Recent results suggest that QAOA depth $p$ has to grow at least logarithmically with problem size to allow quantum advantage~\cite{braviyobstacles,farhi2020qaoaneedsfullgraphtypical,farhi2020qaoaneedsfullgraphworst}. At the same time, implementing a single QAOA layer on fixed qubit architectures such as superconducting processors used in this work requires routing qubits via a $\swapgate$ network, which in general incurs a linear cost if two-qubit gates are used~\cite{ogorman2019swapnetwork}. Combined, these two observations lead to QAOA circuit depth scaling as $O(n\log{n})$ with the number of variables $n$. In the presence of constant error rates, the quantum state fidelity and probability of success decay exponentially with the number of operations if no error correction is performed. In the absence of significant reduction in error rates, this makes it unlikely that circuits with more than logarithmic depth can lead to quantum advantage for optimization~\cite{2009.05532}.

The development of effective error mitigation techniques is therefore a prerequisite for quantum advantage with QAOA. In order to keep the overhead of error mitigation low, the specific structure of the QAOA ansatz must be leveraged. This paper takes the first step towards developing such techniques. 

We introduce an application-specific error mitigation approach for QAOA that leverages the structure of the classical optimization problem to be solved. The proposed methods can be combined with other application-agnostic error mitigation approaches, which we demonstrate by combining  with measurement error mitigation. We show that the proposed symmetry verification procedure is beneficial even on the current hardware, and we demonstrate its promise in lower-error regime by performing numerical simulations. 

Our methods are complementary to the technique proposed in \cite{Streif2021} to reduce the overhead of the bit flip code by leveraging the local particle number conservation (LPNC) property of the ansatz. For example, if a problem exhibits both LPNC and a variable index permutation symmetry, both the symmetry verification and the local particle number verification with the modified bit flip code can be performed in sequence. A promising future direction is applying the techniques developed in this paper to the problem symmetries that arise from the mappings of integer problems onto qubits beyond LPNC.

As higher-fidelity hardware becomes available, we are hopeful that application-specific methods such as the ones introduced in this paper will help bridge the gap between noisy intermediate-scale quantum devices of today and the fault-tolerant quantum computers of tomorrow, unlocking practical quantum advantage along the way.

\section{Acknowledgments}

We thank Stefan Wild, Dmitry Fedorov, Jeffrey Larson, Xiaoyuan Liu, Matthew Otten and Stuart Hadfield for helpful discussions and feedback on the manuscript. We acknowledge the use of IBM Quantum services for this work. In this paper we use \texttt{ibmq\_bogota} and \texttt{ibmq\_jakarta}, which are IBM Quantum Falcon processors. 

\bibliographystyle{myIEEEtran}
\bibliography{references}

\vspace{3em}

\small

\framebox{\parbox{\linewidth}{
The submitted manuscript has been created by UChicago Argonne, LLC, Operator of 
Argonne National Laboratory (``Argonne''). Argonne, a U.S.\ Department of 
Energy Office of Science laboratory, is operated under Contract No.\ 
DE-AC02-06CH11357. 
The U.S.\ Government retains for itself, and others acting on its behalf, a 
paid-up nonexclusive, irrevocable worldwide license in said article to 
reproduce, prepare derivative works, distribute copies to the public, and 
perform publicly and display publicly, by or on behalf of the Government.  The 
Department of Energy will provide public access to these results of federally 
sponsored research in accordance with the DOE Public Access Plan. 
http://energy.gov/downloads/doe-public-access-plan.}}
\end{document}